\documentclass[12pt]{iopart}

\usepackage{graphicx}
\usepackage{dsfont}
\usepackage{iopams}
\usepackage{dsfont}
\usepackage{epstopdf}

\begin{document}

\title{Full-revivals in 2-D Quantum Walks}

\author{M. \v Stefa\v n\'ak$^{(1)}$, B. Koll\'ar$^{(2)}$, T. Kiss$^{(2)}$ and I. Jex$^{(1)}$}
\address{$^{(1)}$ Department of Physics, FJFI \v CVUT v Praze, B\v
rehov\'a 7, 115 19 Praha 1 - Star\'e M\v{e}sto, Czech Republic}
\address{$^{(2)}$ Department of Quantum Optics and Quantum Information, Research
Institute for Solid State Physics and Optics, Hungarian Academy of
Sciences, Konkoly-Thege M. u. 29-33, H-1121 Budapest, Hungary}


\ead{martin.stefanak@fjfi.cvut.cz}

\begin{abstract}
Recurrence of a random walk is described by the P\'olya number. For quantum walks, recurrence is understood as the return of the walker to the origin, rather than the full-revival of its quantum state. Localization for two dimensional quantum walks is known to exist in the sense of non-vanishing probability distribution in the asymptotic limit. We show on the example of the 2-D Grover walk that one can exploit the effect of localization to construct stationary solutions. Moreover, we find full-revivals of a quantum state with a period of two steps. We prove that there cannot be longer cycles for a four-state quantum walk. Stationary states and revivals result from interference which has no counterpart in classical random walks.
\end{abstract}

\section{Introduction}

Quantum walks were proposed \cite{aharonov} as a generalization of classical random walks to quantum mechanical systems. Continuous-time quantum walks \cite{farhi} are suitable for the description of coherent transport in networks \cite{muelken:prl,muelken:pre1} while discrete-time quantum walks \cite{meyer1,meyer2,watrous} have found application in quantum search algorithms \cite{shenvi:2003,kendon:2006,aurel:2007,vasek}. A connection between the two types of quantum walks has been established \cite{strauch,chandra}. Recently, it has been shown that the quantum walk can be regarded as a universal computational primitive \cite{childs}.

One of the interesting problems for random walks is to find the probability of the return to the origin. P\'olya showed \cite{polya} that this probability for balanced classical random walks on regular square lattices equals unity only on the line and the plane. Such random walks are called recurrent. For random walks on higher-dimensional lattices the probability of return to the origin (the P\'olya number) is strictly less than one and such random walks are said to be transient. The concept of recurrence has been extended to quantum walks \cite{prl} and it has been shown that the P\'olya number of a quantum walk is affected by the additional degrees of freedom offered by quantum mechanics \cite{pra} and can differ considerably from the analogous properties of the classical random walks. As pointed out in \cite{chandra2} the recurrence of a quantum walk is understood as a partial revival of the probability at the origin, rather than a full-revival of a quantum state. Nevertheless, we show that under some conditions stationary states and full-revivals are possible for quantum walks. The key feature is localization which is even stronger requirement than recurrence. Indeed, for localizing quantum walks the probability distribution is 
peaked at the origin and, moreover, has a non-vanishing asymptotic limit.

We concentrate on the 2-D Grover walk which has been extensively studied \cite{mackay,tregenna} and which exhibits the unusual effect of localization \cite{inui}. Localization arises from the fact that the propagator of the quantum walk in the momentum representation has eigenvalues which do not depend on the momenta \cite{pra}. Hence, the propagator of the quantum walk in the position representation has a point spectrum and the corresponding eigenvectors are stationary states. Since there are two different eigenvalues the linear combinations of their corresponding eigenvectors results in a simple oscillatory behaviour which shows full quantum state revivals.

The manuscript is organized as follows: we introduce the Grover walk on plane in Section~\ref{sec2}. In Section~\ref{sec3} we discuss the stationary states and full-revivals which emerge in our model. In Section~\ref{sec4} we analyze the point spectrum of the propagator for a general four-state quantum walk on a plane. We conclude by presenting an outlook in Section~\ref{sec5}.

\section{Grover Walk on a Plane}
\label{sec2}

Let us consider the Grover walk on a plane. The Hilbert space of the particle has the form of the tensor product
\begin{equation}
{\cal H} = {\cal H}_P\otimes{\cal H}_C.
\end{equation}
The position space ${\cal H}_P$ is spanned by vectors $|\mathbf{m}\rangle$, where $\mathbf{m}$ is a two-component integer vector corresponding to the location of the particle on the lattice. The particle is allowed to move by one step in all four directions in the plane. To each of these directions we assign a basis vector of the coin space ${\cal H}_C$, i.e.
\begin{equation}
{\cal H}_C = \mathds{C}^4 = \textrm{Span}\left\{|R\rangle,|L\rangle,|U\rangle,|D\rangle\right\}.
\end{equation}
A single step of the quantum walk is given by the propagator
\begin{equation}
U = S \left(I\otimes C\right),
\label{qw:time}
\end{equation}
where $C$ denotes the coin operator and the displacement $S$ has the form
\begin{eqnarray}
\nonumber S & = & \sum\limits_{n,m=-\infty}^{+\infty}\left(\frac{}{}|m + 1, n\rangle\langle m, n|\otimes|R\rangle\langle R|+|m - 1, n\rangle\langle m, n|\otimes|L\rangle\langle L|\right.\\
& & \left.\frac{}{}+ |m , n + 1\rangle\langle m, n|\otimes|U\rangle\langle U| + |m, n-1\rangle\langle m, n|\otimes|D\rangle\langle D|\right).
\end{eqnarray}
As a coin operator we choose the $4\times 4$ Grover matrix
\begin{equation}
C = G = \frac{1}{2}\left(
        \begin{array}{cccc}
          -1 & 1 & 1 & 1 \\
           1 & -1 & 1  & 1 \\
           1 &  1 & -1 & 1 \\
           1 &  1 & 1 & -1 \\
        \end{array}
      \right).
\end{equation}

The state of the particle after $t$ steps is given by the successive application of the propagator $U$ on the initial state. Since the model is translationally invariant the time evolution is greatly simplified in the Fourier picture \cite{pra}. In the momentum representation the time-evolution equation turns into a single difference equation where the key role is played by the propagator
\begin{equation}
\tilde{U}(\mathbf{k}) = \textrm{Diag}\left(e^{i k}, e^{-i k}, e^{i l}, e^{-i l}\right)\cdot G \, .
\label{propagator}
\end{equation}
The time evolution in the momentum representation is solved by diagonalizing the propagator $\tilde{U}(\mathbf{k})$ and by means of the inverse Fourier transformation we can find the solution in the position representation. The behaviour of the quantum walk is thus determined by the spectrum of the propagator $\tilde{U}(\mathbf{k})$ \cite{pra}.

\section{Stationary States and Full-revivals}
\label{sec3}

After specifying our model let us discuss the stationary states and full-revivals. The key feature is that the spectrum of the propagator $\tilde{U}(\mathbf{k})$, expressed in Eq. (\ref{propagator}), contains two eigenvalues independent of the momenta \cite{pra}, namely $\lambda_1 = 1$ and $\lambda_2 = -1$. Hence, the propagator in the position representation $U$, given by Eq. (\ref{qw:time}), has a point spectrum with the same eigenvalues. The corresponding eigenvectors are indeed stationary states. Moreover, we have two different eigenvalues and we can choose the initial state as a linear combination of the eigenvectors. This results in a state which will undergo a cyclic evolution. Since $\lambda_1^2 = \lambda_2^2 = 1$ the length of the cycle is two - the particle will oscillate between two states.

Let us now analyze the stationary states in more detail. The non-normalized eigenvectors of $\tilde{U}(\mathbf{k})$ corresponding to the constant eigenvalues $\lambda_{1,2}=\pm 1$ have the following form
\begin{eqnarray}
\nonumber v_1(\mathbf{k}) & = & \left(e^{i k} (1 + e^{i l}), 1 + e^{i l}, e^{i l} (1 + e^{i k}),
 1 + e^{i k}\right)\\
v_2(\mathbf{k}) & = & \left(e^{i k} (1 - e^{i l}), -1 + e^{i l}, e^{i l} (1 - e^{i k}), -1 +
  e^{i k}\right)\, .
\end{eqnarray}

The stationary states in the position representation are given by the Fourier transformation of these eigenvectors. Since the eigenvectors $v_{1,2}(\mathbf{k})$ depend on the momenta $\mathbf{k}$, the stationary states are delocalized. Indeed, if the particle starts the quantum walk from a single position there cannot be any interference in the first step. Hence, the localized initial state has to spread. Performing the Fourier transformation of the above written eigenvectors and normalizing we find the stationary states
\begin{eqnarray}
\nonumber |\psi_1\rangle & = & \frac{1}{\sqrt{8}}\left(\frac{}{}|0,0\rangle(|L\rangle + |D\rangle) + |1,0\rangle(|R\rangle + |D\rangle) + \right.\\
\nonumber & & \left. \frac{}{} + |0,1\rangle(|L\rangle + |U\rangle) + |1,1\rangle (|R\rangle + |U\rangle) \right)\\
\nonumber |\psi_2\rangle & = & \frac{1}{\sqrt{8}}\left(\frac{}{}-|0,0\rangle(|L\rangle + |D\rangle) + |1,0\rangle(|R\rangle + |D\rangle) + \right.\\
 & & \left. \frac{}{} + |0,1\rangle(|L\rangle + |U\rangle)  -|1,1\rangle (|R\rangle + |U\rangle) \right)\, .
\label{eigen:vec}
\end{eqnarray}
We note that due to the translational invariance of 2-D Grover walk there are infinitely many stationary states which are created by simply shifting the eigenvectors specified in (\ref{eigen:vec}) to a different lattice point. Hence, each of the eigenvalue of the propagator $U$ is infinitely degenerate.

\begin{figure}[h]
\begin{center}
\includegraphics[width=0.3\textwidth]{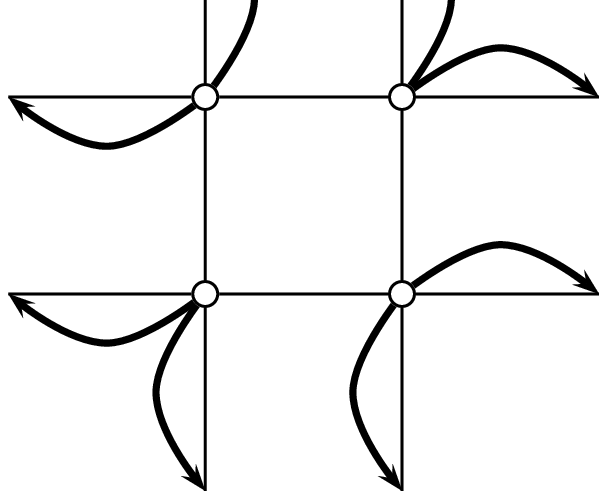}\hfil
\includegraphics[width=0.3\textwidth]{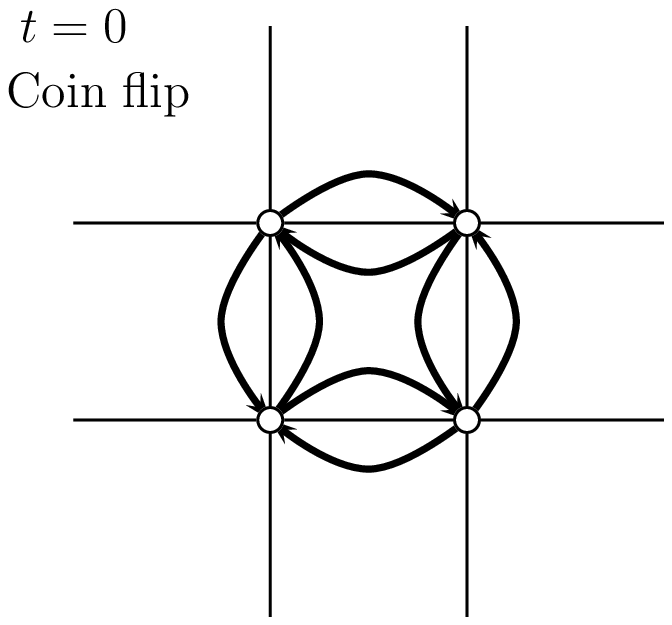}\hfil
\includegraphics[width=0.3\textwidth]{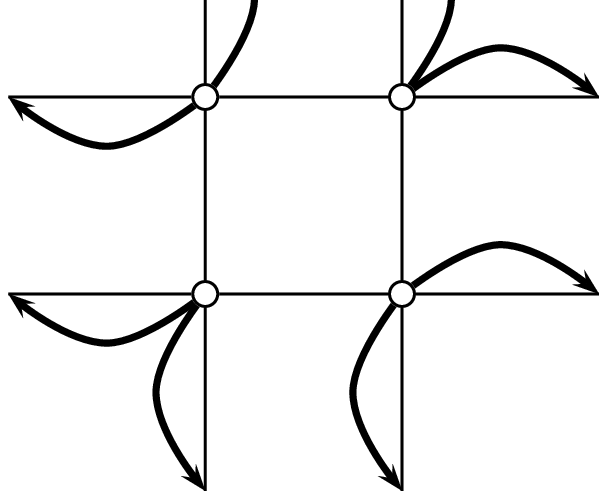}
\caption{The stationary state of the 2-D Grover walk. In the left plot we show the initial state of the quantum walk which is chosen to be $|\psi_1\rangle$. The position of the particle is depicted by the circles and the coin states are denoted by the arrows. As we illustrate in the center plot, the coin flip results in reverting the initial coin states. After the displacement the particle returns to the initial state, as we show in the right plot. Indeed, the state $|\psi_1\rangle$ is an eigenvector of the propagator.}
\label{stat:sol}
\end{center}
\end{figure}

\begin{figure}[h]
\includegraphics[width=0.3\textwidth]{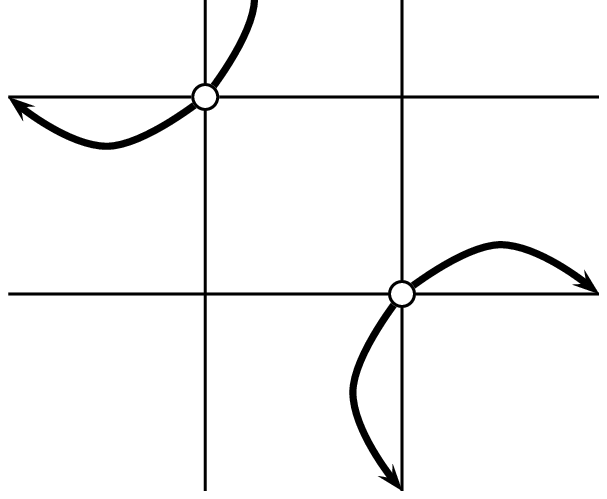}\hspace{11pt}
\includegraphics[width=0.3\textwidth]{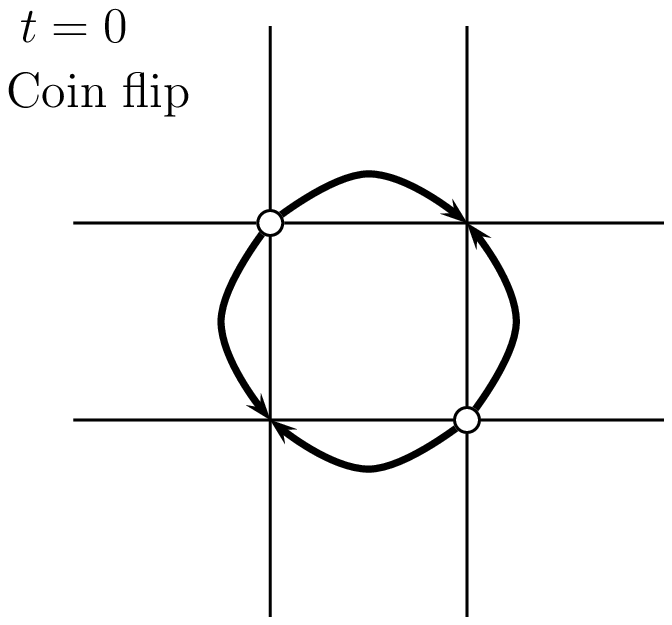}\\
\includegraphics[width=0.3\textwidth]{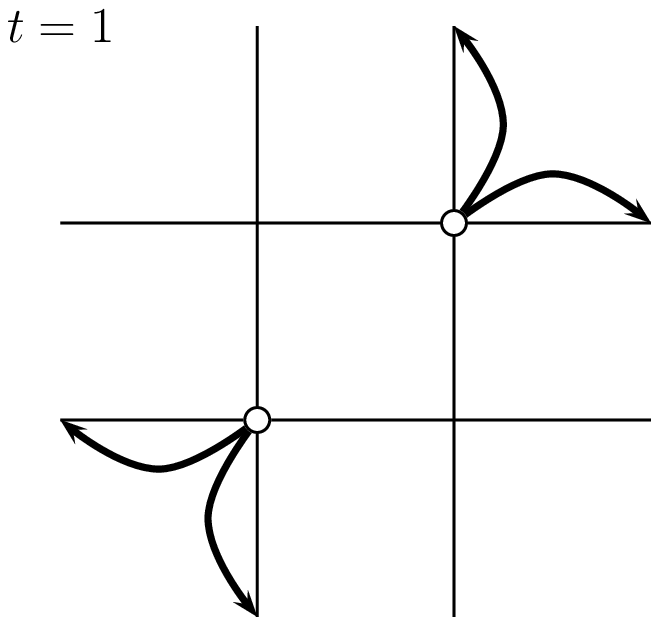}\hfil
\includegraphics[width=0.3\textwidth]{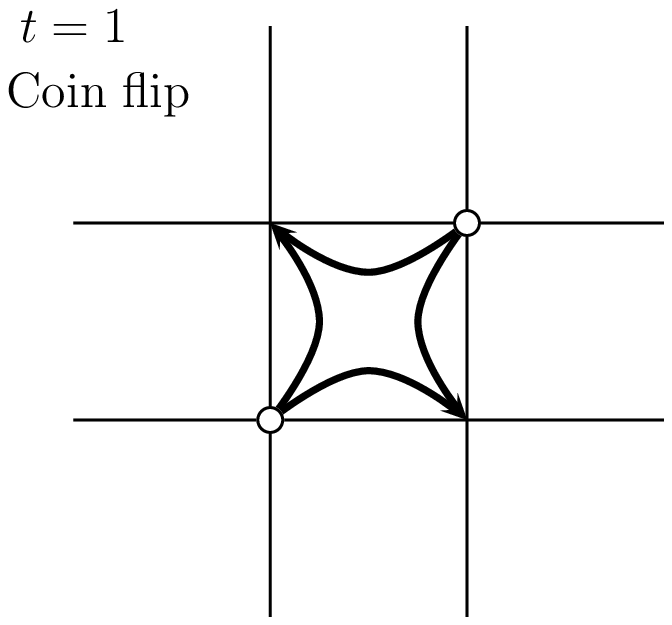}\hfil
\includegraphics[width=0.3\textwidth]{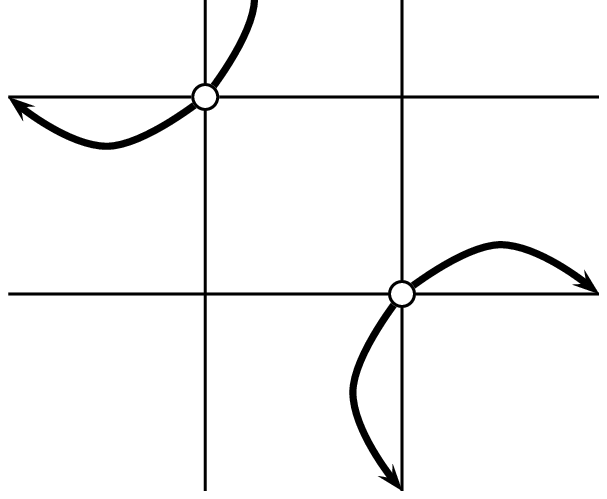}
\caption{Full-revival in the 2-D Grover walk. As the initial state we choose $|\psi(0)\rangle$ given by (\ref{init:rev}) which is not an eigenvector of the evolution operator. Coin flip reverts the initial states of the coin, as can be seen from the upper center plot. The lower left plot indicates that displacing the particle evolves it into an orthogonal state. Indeed, the particle is now located on different lattice points then in the previous step. This is a consequence of the fact that $|\psi(0)\rangle$ is an equal superposition of the two eigenvectors. Finally, applying the coin flip and the displacement returns the particle into the initial state.}
\label{revival}
\end{figure}

We illustrate these results in Figures \ref{stat:sol} and \ref{revival}. In Figure \ref{stat:sol} we show the behaviour of the eigenvector $|\psi_1\rangle$ during a single step of the quantum walk. Let us recall that a single step consists of two parts - first we apply the coin flip and then we conditionally displace the particle. From the form of the eigenvector $|\psi_1\rangle$ given by (\ref{eigen:vec}) we see that at each of the occupied lattice positions the coin state is an equal superposition of two basis states. An important property of the Grover coin is that it turns such a superposition into an equal superposition of the other two basis states, e.g.
\begin{equation}
G (|R\rangle + |U\rangle) = |L\rangle + |D\rangle.
\end{equation}
Hence, the coin flip applied to the eigenvector $|\psi_1\rangle$ results in reverting the initial coin states, as can be seen from the comparison of the left and center plot. Performing the displacement we return the particle into the initial state, as illustrated in the right plot.

In Figure \ref{revival} we choose as the initial state an equal superposition of the two eigenvectors
\begin{equation}
|\psi(0)\rangle = \frac{1}{\sqrt{2}}(|\psi_1\rangle + |\psi_2\rangle) = \frac{1}{2}\left(\frac{}{} |1,0\rangle(|R\rangle + |D\rangle) + |0,1\rangle(|L\rangle + |U\rangle)\right),
\label{init:rev}
\end{equation}
which is not an eigenvector of the propagator. Indeed, we immediately see that
\begin{equation}
|\psi(1)\rangle = \frac{1}{\sqrt{2}}(|\psi_1\rangle - |\psi_2\rangle) = \frac{1}{2}\left(\frac{}{} |0,0\rangle(|L\rangle + |D\rangle) + |1,1\rangle(|R\rangle + |U\rangle) \right),
\end{equation}
which differs from the initial state. Since the initial state is an equal superposition of the two eigenvectors corresponding to different eigenvalues after a single step of the quantum walk it evolves into an orthogonal state. As a consequence of the relation $\lambda_1^2 = \lambda_2^2 = 1$ we find that $|\psi(2)\rangle = |\psi(0)\rangle$. Hence, the state $|\psi(0)\rangle$ undergoes a cyclic evolution with a period of two steps.

Quantum walks showing full-revivals with a period longer than two steps would require either a propagator which has two eigenvalues with a phase difference distinct from $\pi$ or a propagator with more than two eigenvalues. However, in the following section we show that this is not possible for a four-state quantum walk on a plane.

\section{Point Spectrum of a General Propagator}
\label{sec4}

In this section we show that in general the propagator of a four-state quantum walk on a plane has either empty point spectrum or it has two eigenvalues with a $\pi$ phase difference. Let $C$ be an arbitrary $4\times 4$ unitary matrix, the propagator of the quantum walk on a plane is then given by
\begin{equation}
\tilde{U}(k,l) = \textrm{Diag}\left(e^{i k}, e^{-i k}, e^{i l}, e^{-i l}\right)\cdot C.
\end{equation}
Clearly, all four eigenvalues of $\tilde{U}(k,l)$ cannot be independent of the momenta. The case of three constant eigenvalues can be also ruled out. Indeed, the determinant of $\tilde{U}(k,l)$ equals the determinant of $C$, i.e. the product of the eigenvalues of $\tilde{U}(k,l)$ is constant. Hence, if $\tilde{U}(k,l)$ has three constant eigenvalues the fourth eigenvalue has to be constant as well.

Let us now assume that $\lambda$ is a momentum independent eigenvalue of $\tilde{U}(k,l)$. We show that in such a case $-\lambda$ is also an eigenvalue. The characteristic polynomial of $\tilde{U}(k,l)$ has the following form
\begin{equation}
\lambda^4 - a(e^{\pm ik},e^{\pm il}) \lambda^3 + \left(b + c(e^{\pm ik\pm il})\right)\lambda^2 - d(e^{\pm ik},e^{\pm il})\lambda +\det C = 0\, .
\label{char:poly}
\end{equation}
The explicit form of $a,b,c,d$ can be readily calculated but for our purpose only the functional dependence of the coefficients on $k$ and $l$ is of importance. Suppose now that $\lambda$ is a constant solution of (\ref{char:poly}). By taking the derivative of the characteristic polynomial (\ref{char:poly}) with respect to both $k$ and $l$, it is easy to see that $c=0$. Moreover, since the equation (\ref{char:poly}) has to be satisfied for all values of $k$ and $l$,  the following conditions have to be fulfilled independently
\begin{eqnarray}
\nonumber a(e^{\pm ik},e^{\pm il})\lambda^3 + d(e^{\pm ik},e^{\pm il}) \lambda & = & 0\, ,\\
\lambda^4 + b \lambda^2 + \det C & = & 0\, .
\end{eqnarray}
We see that $-\lambda$ satisfies both of these equations. Hence, $-\lambda$ is a solution of the characteristic polynomial (\ref{char:poly}), i.e. $-\lambda$ is an eigenvalue of the propagator in the momentum representation $\tilde{U}(k,l)$.

Due to the connection between the point spectrum of the propagator in the position representation $U$ and the constant eigenvalues of the propagator in the momentum representation $\tilde{U}(k,l)$ we conclude that the point spectrum of $U$ is either empty or equals $\left\{\pm\lambda\right\}$. Since the global phase of the propagator is irrelevant we can choose $\lambda=1$.

\section{Conclusions}
\label{sec5}

We have demonstrated on the example of the 2-D Grover walk that quantum walks can have stationary states or can show full-revivals. The existence of full-revivals in the evolution of a quantum walk is another fascinating effect emphasizing the difference between classical and quantum walks. It is a direct consequence of the non-empty point spectrum of the propagator driving the quantum walk. For the four-state Grover walk on a plane the point spectrum is rather simple which results in two-step revivals. Moreover, we have shown that revivals with longer period than two steps cannot be achieved for a four-state quantum walk. To find quantum walks with a broader point spectrum leading to intriguing revival dynamics is the current scope of our investigation.

\ack

We thank E. Kajari, J. Bergou and S. Pascazio for stimulating discussions. The financial support by MSM 6840770039, M\v SMT LC 06002, the Czech-Hungarian cooperation project (KONTAKT,CZ-10/2007) and by the Hungarian Scientific Research Fund (Contract No. T049234) is gratefully acknowledged.

\end{document}